\documentstyle[PASJadd]{PASJ95}

\begin{document}
\setcounter{page}{000}

\title{The Detection Rate of Molecular Gas in Elliptical Galaxies:
Constraints on Galaxy Formation Theories}

\author{Yutaka {\sc Fujita}\thanks{ Present Address: National
Astronomical Observatory, Mitaka, Tokyo 181-8588} 
$\:$\thanks{JSPS Research Fellow, yfujita@th.nao.ac.jp}, 
and Masahiro {\sc Nagashima}$^{*}$ \\ {\it Department of Earth
and Space Science, Graduate School of Science, Osaka University,}
\\
{\it Machikaneyama-cho,
  Toyonaka, Osaka 560-0043}
\\[6pt]
and \\
Naoteru {\sc Gouda} \\
{\it National Astronomical Observatory, Mitaka, Tokyo 181-8588}}

\abst{In order to constrain parameters in galaxy formation theories,
especially those for a star formation process, we investigate cold gas
in elliptical galaxies.  We calculate the detection rate of cold gas in
them using a semi-analytic model of galaxy formation and compare it with
observations.  We show that the model with a long star formation
time-scale ($\sim 20$ Gyr) is inconsistent with observations.  Thus,
some mechanisms of reducing the mass of interstellar medium, such as the
consumption of molecular gas by star formation and/or reheating from
supernovae, are certainly effective in galaxies.  Our model predicts
that star formation induced when galaxies in a halo collide each other
reduces the cold gas left until the present.  However, we find that the
reduction through random collisions of satellite (non-central) galaxies
in mean free time-scale in a halo is not required to explain the
observations. This may imply that the collisions and mergers between
satellite galaxies do not occur so often in clusters or that they do not
stimulate the star formation activity as much as the simple collision
model we adopted.  For cD galaxies, the predicted detection rate of cold
gas is consistent with observations as long as the transformation of hot
gas into cold gas is prevented in halos whose circular velocities are
larger than $500\;\rm km\; s^{-1}$.  Moreover, we find that the cold gas
brought into cDs through captures of gas-rich galaxies is little.  We
also show that the fraction of galaxies with observable cold gas should
be small for cluster ellipticals in comparison with that for field
ellipticals.  Our fiducial models and the models with large reheating
efficiency can reproduce observations well, although the comparison with
a larger and complete sample of elliptical galaxies will constrain
physical parameters in galaxy formation theories more strictly.}

\kword{Galaxies: clusters: general --- Galaxies: ISM --- Galaxies:
cooling flows --- Galaxies: elliptical and lenticular, cD}

\maketitle
\thispagestyle{headings}

\section{Introduction}
\label{sec:intro}

Theoretical models based on hierarchical clustering scenario are used to
show the merging history and evolution of galaxies with various masses
and of species simultaneously. N-body simulations with hydrodynamics
have been often used to predict them. Although these simulations
reproduce an outline of the formation and evolution of galaxies, it is
not suitable for detailed study of baryonic component of galaxies at
present. This is because many processes such as gas cooling, star
formation, and supernova feedback are responsible for the evolution of
the baryonic component and a large dynamic range in simulation is needed
to treat those processes.

Semi-analytic models of galaxy formation are embedded within the
framework of the hierarchical clustering scenario. In these models, gas
cooling, star formation, and supernova feedback are approximated by
simple rules. Contrary to N-body simulations, they can simultaneously
treat physical processes in various scales such as formation of large
scale structure in the universe and star formation in a galaxy. However,
although the semi-analytic models have achieved notable successes in
modeling many properties of galaxies, such as a luminosity function of
galaxies (Kauffmann et al. 1993; Cole et al. 1994; Kauffmann, Charlot
1998; Baugh et al. 1998; Somerville, Primack 1999; Nagashima et
al. 1999), the models still have many unknown parameters. In particular,
the parameters regarding a star formation process such as the efficiency
of supernova feedback and the time-scale of star formation have not been
determined, although they are key parameters of galaxy formation. Since
it is difficult to derive them only from theoretical studies at present,
we need independent observational constraints to determine
them. Observations of luminosities and colors of galaxies, or
information on stars, have mainly been used as the constraints, but they
appear to be insufficient. Thus we need other constraints.

In this paper, we investigate the evolution of cold gas in elliptical
galaxies using a semi-analytic model, and compare the results with
observations. Although the cold gas in general galaxies has been
investigated (e.g. Somerville, Primack 1999), this is the first time to
constrain the parameters regarding a star formation process by focusing
on cold gas in elliptical galaxies. In the following, we classify
elliptical galaxies into cD galaxies (cDs), cluster elliptical galaxies
(clEs), and field elliptical galaxies (fiEs).

cD galaxies are the ones which dominate centers of galaxy clusters. So
far cold gas in cD galaxies has been investigated from the viewpoint of
cooling flows. In the absence of heating, the intracluster gas around cD
galaxies is inferred to be cooling at a rate of $\dot{M}_{\rm CF}\sim
100\MO\rm\; yr^{-1}$ (Fabian 1994). Thus, the total mass accumulated
around the cD galaxies would result in $\sim 10^{12}\MO$ if the cooling
occurred steadily at the rate over the Hubble time. However, such a
large amount of cold gas has not been detected in cD galaxies
(e.g. Burns et al. 1981; Valentijn, Giovanelli 1982; McNamara et
al. 1990; McNamara, Jaffe 1994; O'Dea et al. 1994; Fujita et al. 2000)
except for NGC1275 in the Perseus cluster (e.g. Lazareff et al. 1989;
Mirabel et al. 1989; Inoue et al. 1996). On the other hand, the
semi-analytic models predict that cD galaxies have experienced
mergers. Thus, the capture of gas-rich galaxies is another possible
supply route of cold gas into cD galaxies. Fujita et al. (2000) show
that the cold gas brought into cD galaxies should not be evaporated by
the ambient hot intracluster medium (ICM) gas if heat conduction rate is
small as suggested by X-ray observations. Thus, the observations of cold
gas in cD galaxies give the upper limit of the gas supplied through
galaxy capture.

Contrary to cD galaxies, cold gas has been detected in many clEs and
fiEs, although the mass is small ($\sim 10^8 \MO$).  Wiklind et
al. (1995) and Knapp and Rupen (1996) show that more than 40\% of clEs
and fiEs they investigated have CO emission or absorption. Thus, it is
interesting to investigate if the difference of the detection rates of
cold gas between cDs and non-cDs can be explained by the difference of
their evolution histories. Previous studies suggest that some of clEs
and fiEs are expected to have weak cooling flows of $\dot{M}_{\rm
CF}\sim 1\MO\rm\; yr^{-1}$ (e,g, Thomas et al. 1986). However, Wiklind
et al. (1995) indicates that the the cold gas detected has an external
origin.

The plan of our paper is as follows. The following section~2 describes
the our model based on hierarchical clustering scenario. In \S3, we 
predict the cold gas in elliptical galaxies and compare it with
observations. We present our conclusions in \S4.

\section{Models}
\label{sec:dis}

We use the semi-analytic model of Nagashima, Gouda 1999 (hereafter NG).
The model includes some physical processes connected with galaxy
formation, such as merging histories of dark halos, gas cooling, star
formation, supernova feedback, mergers of galaxies, and so on.  In order
to realize the various merging paths of galaxies, the model is based on
the Monte Carlo method.  The outline of the procedure is as follows.
First, the merging paths of dark halos are realized by the extension of
the Press-Schechter formalism (Press, Schechter 1974).  These paths are
determined by properties of the initial density fluctuations.  In this
paper, we assume the cold dark matter model.  Next, in each step of a
merging path, that is, in each dark halo, evolutions of the baryonic
component, namely, gas cooling, star formation, and supernova feedback,
are calculated. We estimate the mass of cold gas considering cooling
processes.
The stars are formed from the cold gas with a time-scale proportional to
a dynamical time-scale. In the Einstein-de Sitter universe, it is
$\tau_{*}^{0}(1+z)^{-3/2}$, where $z$ is the redshift and $\tau_{*}^{0}$
is the parameter specified below. In other universes, the redshift
dependence (-3/2) changes a little. The cold gas is reheated by
supernovae. The amount of the reheated gas is given by $\beta(V_{\rm
c})\Delta M_{*}$, where $V_{\rm c}$ is the circular velocity, $\Delta
M_{*}$ is the mass of the newly formed stars, and $\beta(V_{\rm c})$ is
the feedback strength, $\beta=(V_{\rm c}/V_{\rm hot})^{-\alpha_{\rm
hot}}$. The two parameters $V_{\rm hot}$ and $\alpha_{\rm hot}$ are
specified below. We recognize a system consisting of the stars and cold
gas as a {\it galaxy}.
As Blumenthal et al. (1984) indicated, the baryonic gas can cool and
galaxies form in dark halos with mass of $\ltsim 10^{12}\MO$ because the
cooling time is smaller than the dynamical time of the halos. Thus, in
these halos at least one galaxy exists. Since our calculations show that
dark halos with mass of $\gtsim 10^{12}\MO$ form through mergers of dark
halos with mass of $\ltsim 10^{12}\MO$, every dark halo has at least one
galaxy. 

When two or more dark halos merge together, there is a possibility that
galaxies contained in progenitor halos merge together.  We consider
mergers between central and satellite galaxies for all models. We define
the central galaxy in the most massive progenitor halo as the central
galaxy in the new halo (see NG; Somerville, Primack 1999); generally, a
central galaxy is the most massive galaxy in the halo. The rests are
defined as satellite galaxies. We assume that radiatively cooled gas in
a dark halo accretes only to the central galaxy. All the satellites lose
their kinetic energy owing to dynamical friction against the dark matter
background, fall in towards the central galaxy, and finally merge with
the central galaxy in a certain time-scale. The judgment whether the
galaxies merge together or not is determined by the dynamical friction
time-scale (equation [7-26] in Binney, Tremaine 1987).
If the dynamical friction time-scale is shorter than the lifetime of the
merged new halo, which is defined as the elapsed time from the time when
the halo forms through the merger of two or more dark halos with
comparable masses to the time when the halo experiences another merger
with halos with comparable masses, the galaxies merge. (`Comparable'
means that the mass ratio of the largest halo to others is smaller than
five.) If not, the galaxies do not merge and the common dark halo has
two or more galaxies. This system is recognized as a group or cluster of
galaxies. We assume that when galaxies with comparable masses merge,
where `comparable' means that their mass ratio is smaller than five,
their cold gas turns into stars simultaneously (starburst). Finally, we
calculate the color and luminosity of each galaxy from its star
formation history by using a population synthesis model.  In this paper,
we use the population synthesis code given by Kodama and Arimoto (1997). 
Through the above procedure, the properties of each model galaxy can be
calculated. More details on our model are found in NG.

Most of the parameters we use here are fixed at those in NG;
cosmological parameters are $\Lambda=0$ and $\Omega_b=0.06$. We consider
the Einstein-de Sitter universe ($\Omega_0=1$, $h=0.5$, $\sigma_8=0.67$)
and an open universe ($\Omega_0=0.3$, $h=0.7$, $\sigma_8=1$), where
$H_0=100\;h\;\rm km\; s^{-1}\;Mpc^{-1}$. The feedback parameter in NG is
fixed at $\alpha_{\rm hot}=2$, which means that the mass fraction
reheated by supernovae is proportional to the inverse of the depth of
the potential well of the halos. In addition to $\Omega_0$, we vary two
parameters, the circular velocity of a galaxy under which reheating from
supernovae becomes effective ($V_{\rm hot}$) and the time-scale of star
formation ($\tau_{\star}^0$). We ignore the evaporation of interstellar
medium (ISM) by hot ICM because observations show that the heat
conduction rate of ICM is small enough (Fujita et al. 2000). We also
ignore ram-pressure stripping of ISM by ICM for simplicity. However, it
is to be noted that the feedback parameter $V_{\rm hot}$ can implicitly
include the effects of evaporation and stripping for galaxies in
clusters even if they are effective. If they are effective, star
formation in galaxies is prevented, which is the same as the case of
strong feedback.

We list the parameters in table~1. Models A and A' are fiducial models
and the predictions of the models are consistent with observations such
as a luminosity function of galaxies. Models B and B' are the extreme
cases where cold gas in galaxies easily accumulates. That is, the star
formation time-scale is very long and the cold gas is not much consumed. 
Models C and C' are the opposites of models B and B'; the reheating is
effective and cold gas hardly accumulates except for extremely massive
galaxies. Models D and D' are the same as models A and A', respectively,
except for including the effect of star formation when satellite
galaxies in a dark halo collide and merge with each other; in other
models we do not consider the collisions between satellites. We
emphasize that the collisions between satellites are random ones, which
are distinct from collisions between central galaxies and satellite
galaxies induced by dynamical friction. We adopt the collision and
merger model of Somerville and Primack (1999; see also NG) that is based
on the collision time-scale derived by Makino and Hut (1997) using
numerical simulations. Following the model of Somerville and Primack
(1999), we do not allow random collisions between central galaxies and
satellite galaxies. As is the case of the collision between central
galaxies and satellite galaxies, a starburst occurs when satellites with
comparable masses merge together.

We define an elliptical galaxy as the one whose B-band bulge to total
luminosity ratio is larger than 0.6. Among the elliptical galaxies, cDs
and clEs reside in dark halos with circular velocities of $V_{\rm c} =
1000$ and $600\rm\;km \;s^{-1}$; the halos with $V_{\rm c} = 1000$ and
$600\rm\;km \;s^{-1}$ correspond to a cluster of galaxies and a group of
galaxies, respectively. cDs are the most massive galaxies in the dark
halos and clEs are the rests.  The results in the following show that
these `cDs' are at least a few times as massive as other galaxies in the
halos. In most cases, cDs are central galaxies. Field ellipticals (fiEs)
are the elliptical galaxies that reside in the dark halos of $V_{\rm c}
= 220\rm\;km \;s^{-1}$.

\section{Results and Discussion}

\begin{table}
\small
\begin{verse}
Table~1.\hspace{4pt}Model parameters.\\ 
\end{verse}
\vspace{6pt}
\begin{tabular*}{\columnwidth}{@{\hspace{\tabcolsep}
\extracolsep{\fill}}ccccc}
\hline\hline\\[-6pt]
 Model & $V_{\rm hot}\;(\rm km\; s^{-1})$ & $\tau_{\star}^0$ (Gyr)&
$\Omega_0$& collision \\
[4pt]\hline\\[-6pt]
 A   &   140   &   5   &1   & no \\
 B   &   140   &   20  &1   & no \\
 C   &   280   &   5   &1   & no \\
 D   &   140   &   5   &1   & yes \\
 A'  &   140   &   5   &0.3 & no \\
 B'  &   140   &   20  &0.3 & no \\
 C'  &   280   &   5   &0.3 & no \\
 D'  &   140   &   5   &0.3 & yes \\[4pt]
\hline
\end{tabular*}
\end{table}

\begin{table*}[t]
\small
\begin{center}
Table~2.\hspace{4pt}Molecular gas detection rate.\\ 
\end{center}
\vspace{6pt}
\begin{tabular*}{\textwidth}{@{\hspace{\tabcolsep}
\extracolsep{\fill}}cccccc}
\hline\hline\\[-6pt]
 Model & cD & clE & cD & clE & fiE \\
  &$(V_{\rm c}=1000)$&$(V_{\rm c}=1000)$&
$(V_{\rm c}=600)$&$(V_{\rm c}=600)$&
\\
[4pt]\hline\\[-6pt]
A &0.02(1/47)&0.65(1098/1684)&0.18(33/181)&0.69(737/1067)&0.66(754/1146)\\
B &0.06(3/47)&0.72(893/1238) &0.31(57/185)&0.77(622/812)&0.85(1457/1714)\\
C &0.00(0/47)&0.42(442/1062) &0.14(24/173)&0.46(292/641) &0.59(343/581)\\
D &0.02(1/45)&0.09(526/6132) &0.11(19/180)&0.09(455/4964)&0.32(851/2642)\\
A'&0.10(5/48)&0.28(1288/4641)&0.22(42/193)&0.24(848/3498)&0.53(850/1599)\\
B'&0.10(5/48)&0.78(3319/4247)&0.36(70/197)&0.82(2692/3293)&0.90(1977/2191)\\
C'&0.08(4/48)&0.26(527/2062) &0.21(41/192)&0.17(239/1047)&0.58(539/934)\\
D'&0.02(1/50)&0.05(508/10756)&0.11(21/191)&0.05(425/8878)&0.13(1123/8385)\\
[4pt]
\hline
\end{tabular*}
\vspace{6pt}\par\noindent
Numbers of galaxies are presented in parentheses; the numerators are the
 numbers of galaxies with observable cold gas and the denominators are
 the total numbers of galaxies. The units of $V_{\rm c}$ are $\rm km\;
 s^{-1}$.
\end{table*}

The results of Monte Carlo realizations of merging history are shown in
table~2. For each kind of galaxies at $z=0$, we present the predicted
detection rates of elliptical galaxies with cold gas more abundant than
the typical observational limit $M_{\rm cold}$. We take $M_{\rm
cold}=10^9\;\MO$ for cDs and $M_{\rm cold}=10^8\;\MO$ for clEs and fiEs,
because the mean redshift of cDs for which the CO observations have been
made is larger than that of clEs and fiEs. Table~2 shows that the
differences between the predictions of the models of $\Omega_0=1$ and
$\Omega_0=0.3$ are small for cDs and fiEs. In an open universe, the
formation epoch of galaxies and clusters is earlier in comparison with
the Einstein-de Sitter universe. Thus, more ISM is consumed by star
formation. On the contrary, since the time-scale of dynamical friction
in a given halo does not depend on cosmological parameters, the cDs in
an open universe have more chances to capture gas-rich satellite galaxies
and accumulate more cold gas. This is also the case for most fiEs
because $\gtsim 70$\% of fiEs are the central galaxies in the halos
($V_{\rm c}=220\;\rm km\; s^{-1}$). Since these two effects are
canceled, the results for $\Omega_0=1$ and $\Omega_0=0.3$ are not so
much different.  On the other hand, clEs do not capture other gas-rich
galaxies except for models D and D'. Thus, the gas detection rates in an
open universe are lower than those in the Einstein-de Sitter universe. The
low gas detection rates in models D and D' are due to star formation and
gas consumption when satellites collide each other. Moreover, these
reduce the gas supply to central galaxies through the capture of the
satellites.

So far 26 cDs have been searched in CO (those in table 4 in O'Dea et al. 
1994, NGC 1275 and NGC 1129) and it resulted in one detection. On the
contrary, CO emission has been observed in many elliptical galaxies
other than cDs. For example, Wiklind et al. (1995) detected CO in 16
elliptical galaxies out of 29.  Knapp and Rupen (1996) searched for CO
emission from 42 early-type galaxies with $100\rm\;\mu m$ flux densities
greater than 1 Jy. They detected emission from 11 of the galaxies. On
the contrary, although Braine et al. (1997) observed 6 X-ray bright
elliptical galaxies, none of the galaxies were detected in CO.

However, the samples of elliptical galaxies are not complete. Since the
data sets of Wiklind et al. (1995) and Knapp and Rupen (1996) are biased
toward ellipticals known to contain a FIR component, the galaxies do
not represent all elliptical galaxies. Thus, the quantitative comparison
between the observations and our calculations may be difficult at
present, although the samples would increase in the future. Bearing that
in mind, we will discuss the detection rate of CO.

Table~2 shows that the detection rates of CO for clEs and fiEs predicted
by models B and B' are high. Wiklind et al. (1995) detected CO emission
from 67\% of fiEs and from 41\% of clEs. Knapp and Rupen (1996) found
that the detection rate of CO for non-cD ellipticals is 45\% for their
sample and those in literatures, although they do not discriminate
between clEs and fiEs. Quantitatively, the high detection rates in
models B and B' appear to be inconsistent with the
observations. Although we do not present the result, the model with
$V_{\rm c}< 140\;\rm km\; s^{-1}$ and $\tau_{\star}^0=5$ Gyr, that is,
reheating from supernova is ineffective, gives a similar result. These
mean that the consumption of molecular gas by star formation and/or
reheating cannot be ignored in galaxy formation theories. Since
ram-pressure stripping and evaporation by hot ICM do not significantly
affect fiEs, they will not solve the inconsistency for fiEs . For clEs,
however, ram-pressure stripping and evaporation may be effective and may
make the detection rates of cold gas smaller. 

Low detection rates of CO predicted by models D and D' are also
inconsistent with the observations of non-cD galaxies.  This indicates
that our collision-induced starburst model overestimates the gas
consumption when satellites collide each other. This may mean that the
collisions and mergers between satellites seldom occur in a cluster halo
or that they do not stimulate the star formation activity, in comparison
with the prediction of the simple model.

In the following, we discuss the remaining models A, A', C, and C'.
Considering the fact that most of the cD galaxies searched for CO are in
clusters with $V_{\rm c}>700\rm\; km\; s^{-1}$, the detection rates
predicted by these models (especially, models A and C) in table~2 seem
to be consistent with the observations (1/26). There are two reasons for
the lack of cold gas in cDs. One is the assumption that the
transformation of hot gas into cold gas is suppressed after the circular
velocities of the halos grow up to $V_{\rm c}=500\;\rm km\;
s^{-1}$. Since the transformation stopped more recently for halos with
$V_{\rm c}=600\;\rm km\; s^{-1}$ than for those with $V_{\rm
c}=1000\;\rm km\; s^{-1}$, the detection rates of cold gas in cDs in the
formers are higher. The assumption is often adopted in the field of the
semi-analytic approach to suppress the formation of monster
galaxies. The calculations without the assumption result in large number
density of the monsters, which is inconsistent with the observations
(e.g. Kauffmann et al. 1993). Although the assumption is ad hoc, it may
be justified as follows. In a halo with $V_{\rm c}\gtsim 500\;\rm km\;
s^{-1}$, the cooling time of gas is larger than the dynamical time of
the halo but still smaller than the lifetime of the halo, that is, a
cooling flow is expected to occur (Fabian 1994). As many authors
indicate, cooled gas expected by the cooling flow model becomes neither
normal stars nor cold gas (e.g. Fabian 1994; McNamara, Jaffe 1994;
Fujita et al. 2000). Thus, it may become `baryonic dark matter' such as
dust (Fabian et al. 1994; Voit, Donahue 1995; Edge et al. 1999; Allen
2000) or low mass stars (Sarazin, O'Connell 1983; Mathews, Brighenti
1999). Recently, Wu et al. (1999) consider cooling flows in a
semi-analytic model and indicate that cooled gas should turn to be
`baryonic dark matter' in halos with $\gtsim 10^{12}\MO$; it may not be
detected as cold gas or normal stars. The other reason for the lack of
cold gas in cDs is that the amount of the cold gas brought through
galaxy captures is small. Although the cDs capture four galaxies in
average for $z<1$, the cold gas brought by the galaxies is consumed by
star formation activity and is not left.

Table~2 also shows that the fraction of galaxies with cold gas is small
for clEs in comparison with that for fiEs. This is because clEs are
satellites which do not capture other galaxies (except for models D and
D') while most fiEs are central galaxies which often capture gas-rich
galaxies. Moreover, cooled gas in a dark halo accretes only to the
central galaxy. The tendency is consistent with the observation of
Wiklind et al. (1995) (41\% for clEs and 67\% for fiEs). Quantitatively,
models A and C seem to be preferable to models A' and C'. If
ram-pressure stripping and evaporation of cold gas are effective for
clEs, the remaining cold gas should decrease and inconsistency between
the observations and the predictions of models A' and C' should be more
prominent. However, we think that we need more observations and more
improved models to obtain a definite conclusion about models A, A', C,
and C'.


\section{Conclusions}

In order to constrain the parameters regarding star formation process in
galaxy formation theory, we have investigated the detection rate of cold
gas in elliptical galaxies. Using a semi-analytic model of galaxy
formation, we have predicted the amount of cold gas in elliptical
galaxies and compare it with observations. The models with a large
time-scale of star formation predict high detection rates of cold
gas. They are inconsistent with the observations of non-cD elliptical
galaxies and are quantitatively rejected. This means that the mechanisms
of reducing the mass of the ISM, such as the consumption of molecular
gas by star formation and/or reheating from supernovae, cannot be
ignored in galaxy formation theories. We found that the collisions and
mergers between satellite galaxies in a cluster halo reduce cold gas in
them. However, observations show that the simple collision model we
adopted overestimates the effect, and that the models without the
collisions are preferable. This may imply that the satellites collide
each other less frequently or that the collisions cause star formation
less effectively than the model predicts. For cD galaxies, the predicted
detection rate of cold gas is consistent with the observations as long
as the transformation of hot gas into cold gas is prevented in halos
with $V_{\rm c}\gtsim 500\;\rm km\; s^{-1}$. Moreover, we have shown
that the amount of cold gas brought into cDs through captures of
gas-rich galaxies is small. For elliptical galaxies other than cDs, our
models predict that the fraction of galaxies with observable cold gas is
small for cluster ellipticals in comparison with that for field
ellipticals. In this study, we found that our fiducial models and the
models with large reheating efficiency are consistent with the
observations of cold gas in elliptical galaxies of different types. With
larger samples of elliptical galaxies, the parameters in theories of
galaxy formation will more strictly be constrained.

\par
\vspace{1pc}\par
We thank an anonymous referee for invaluable advice and suggestions. We
also thank for T. Tosaki, A. Nakamichi, and N. Kuno for useful
discussions.  This work was supported in part by the JSPS Research
Fellowship for Young Scientists. Numerical computation was carried out
in part at the Yukawa Institute Computer Facility.

\section*{References}
\small

\re Allen S.W.\ 2000, MNRAS in press

\re Baugh C.M., Cole S., Frenk C.S., Lacey C.G.\ 1998, ApJ 498, 504


\re Blumenthal G.R., Faber s.M., Primack J.R., Rees M.J.\ 1984, Nature
311 517

\re Binney J., Tremaine S.\ 1987, Galactic Dynamics (Princeton;
Princeton)

\re Braine J., Henkel C., Wiklind T.\ 1997, A\&A 321, 765

\re Burns J.O., White R.A., Haynes M.P.\ 1981, AJ 86, 1120


\re Edge A.C., Ivison R.J., Smail I., Blain A.W.,Kneib J.-P.\ 1999,
MNRAS 306, 599


\re Fabian A.C.\ 1994, ARA\&A 32, 77

\re Fujita Y., Tosaki T., Nakamichi A., Kuno N.\ 2000, PASJ, 52, 235


%



\re Inoue M.Y., Kamono S., Kawabe R., Inoue M., Hasegawa T., Tanaka M.\
1996, AJ 1111, 1852

\re Kauffmann G., Charlot S.\ 1998, MNRAS 294, 705

\re Kauffmann G., White S.D.M., Guiderdoni M.\ 1993, MNRAS 264, 201

\re Knapp G.R., Rupen M.P.\ 1996, ApJ 460, 271

\re Kodama T., Arimoto N.\ 1997, A\&A 320, 41

\re Lazareff B., Castets A., Kim D.W., Jura M.\ 1989, ApJ 336, L13

\re Makino J., Hut P.\ 1997, ApJ 481, 83

\re Mathews W.G., Brighenti F.\ 1999, ApJ 526, 114


\re McNamara B.R., Bregman J.N., O'Connell R.W.\ 1990, ApJ 360, 20

\re McNamara B.R., Jaffe W.\ 1994, A\&A 281, 673


\re Mirabel I.F., Sanders D.B., Kazes I.\ 1989, ApJ 340, L9

\re Nagashima M., Gouda N.\ 1999, submitted to MNRAS (NG,
astro-ph/9906184)

\re Nagashima M., Gouda N., Sugiura N.\ 1999, MNRAS 305, 449


\re O'Dea C.P., Baum S.A., Maloney P.R., Tacconi L.J., Sparks W.B.\
1994, ApJ 422, 467


\re Somerville R.S., Primack J.R.\ 1999, MNRAS 310, 1087


\re Press W., Schechter P.\ 1974, ApJ 187, 425


\re Sarazin C.L., O'Connell R.W.\ 1983, ApJ 268, 552




\re Valentijn E.A., Giovanelli R.\ 1982, A\&A, 114, 208

\re Voit G.M., Donahue M.\ 1995, ApJ, 452, 164 


\re Wiklind T., Combes F., Henkel C.\ 1995, A\&A 297, 643

\re Wu K.K.S., Fabian A.C., Nulsen P.E.J. submitted to MNRAS 
(astro-ph/9907112)


\end{document}